\documentclass{elsart}
\usepackage{graphicx}

\begin{document}

\title{Temperature evolution of the structure of liquid $3d$ transition metals: MD study}

\author{A.~Kimmel, I.~Gusenkov}

\address{ Physical Technical Institute of the Ural branch of the
Russian Academy of Sciences, 132, Kirov St., Izhevsk, Russia,
426001}
\ead {las@pti.udm.ru}

\begin{abstract}
The effective pair potentials of liquid $3d$ transition metals
have been derived from the experimental diffraction data by means
of the inverse self-consistent method. The obtained potentials
provide highly accurate coincidence of the simulated by means of
the molecular dynamics method and the experimental structural
factors at the range of temperatures. The statistical analysis of
Voronoy polyhedra of the simulated systems has indicated the
presence of the temperature range of inhomogeneity, within which
the metallic liquids contain several competitive types of the
local order, and out of them structure of liquid becomes homogeneous.
\end{abstract}
\maketitle



\section{INTRODUCTION}
\hfil

One of the major problems in the analysis of diffraction data of
disordered systems is the lack of any general method for
developing structural models that agree quantitatively with the
experimental data. Most analyses are mainly qualitative and are
based on a few data features, such as peak positions and
coordination numbers derived from radial distribution functions.

Sometimes the results of simulations are in good agreement with
experiment (though comparison is made with the radial distribution
functions rather than the structure factors), but usually the
agreement is only qualitative, and occasionally there are major
differences, especially in the descriptions of the liquid state of
the $3d$ transition metals. Usually, the disagreement between
model and experimental results is due to the choice of interactive
potentials [1-6].
In this case, the inverse methods [7-12], in which
the effective pair potentials are derived from the experimental
structural data, is one of the best methods to compute models,
structural properties of which quantitatively agree with the
experimental data. It is also well recognized that the effective
pair potentials estimated from the experimental structural data
are quite useful, because they are considered to include, more or
less, the particular features of liquid of interest.

The other problem of the simulation the disordered system is the
analysis of the obtained information, which represent the complete
list of coordinates and velocities of the atoms. Such information
allows us to describe the realization of the current state of the
system completely. One of the methods of solution of such problems
is the statistical analysis of the Voronoy polyhedra built for
each atom of the system [13,14]
The analysis makes possible the description of
the local order of the simulated system.

In the present work the effective pair potentials for liquid $3d$
transition metals $Fe$, $Co$ and $Ni$ were derived  at the range
of temperatures by means of the inverse method proposed in
\cite{Reatto} from the experimental structural factors
\cite{Waseda}. To test the potentials, the dynamic properties,
calculated by means of the molecular dynamics (MD) method, have
been compared with their experimental values. Finally, the
statistical analysis of Voronoy polyhedra, built on the MD
configurations of the investigated metals, has been carried out.

\section{METHOD}

\subsection{THE SELF-CONSISTENT METHOD}
\hfil

The iterative self-consistent method originally proposed by Reatto
$et~al.$ \cite{Reatto} is based on the computer simulations and a
formally exact result of the integral equations theory, which
relates the effective pair potential with the pair and direct
correlation functions $g(r)$ and $c(r)$, respectively:
\begin{equation}\label{1}
\beta \phi(r) = g(r) - c(r) - 1 - \ln g(r) + B(r),
\end{equation}
where $\beta=1/k_bT$, $B(r)$ is so-called $bridge$-function. The
functions $g(r)$ and $c(r)$ can be easily obtained from the
experimental structural factor $S_{exp}(q)$. However, the
$bridge$-function $B(r)$ can not be obtained the same way. This is
why in the original Reatto method the $bridge$-function of the
hard sphere system $B_{HS}(r,\eta)$, where the external parameter
$\eta$ is the packing factor, was used as an initial estimation
for the $bridge$-function, and potential $\varphi_0(r)$
correspondently. As was pointed in the work \cite{Reatto} the
convergence of the iterative method depends on the initial
approximation of the potential $\varphi_0(r)$. To avoid the
dependence on any external parameters in the presented work the
mean-field potential $\varphi_{MF}(r)$, which characterizes
certain self-consistent field of particles, where the force acting
on the particle of the system is the function of all particles
arrangement, was used as the initial potential:
$$
\varphi_{MF}(r)=-ln (g(r)).
$$

The procedure of deriving effective potential, $\varphi(r)$, is
iterative. The equilibrium atomic configuration simulated by means
of the molecular dynamics method with given potential
$\varphi_{i}(r)$ corresponds to the pair correlative function
$g_i(r)$, direct correlative functions, $c_i(r)$ and structural
factor, $S_i(q)$. The next iteration constructs the corrected
potential:

$$
\beta\varphi(r)_{i+1}=g_{exp}(r)-1-c_{exp}(r)-\ln{g_{exp}(r)}+B_i(r),
$$
where the $bridge$-function $B_i(r)$ exactly corresponds to the
potential $\varphi_i(r)$:

$$
B_i(r)=\beta\varphi_i(r) - g_i(r)+c_i(r)+\ln{g_i(r)}.
$$

The iterative process is repeated until the discrepancy ~$\chi$
between the criterion (experimental) and simulated functions
becomes smaller than the desired accuracy:

\begin{equation}
\chi=\frac{1}{N}\sum_{n}^{N}[g_n^{exp}(r)-g_n^\tau(r)]^2
+\frac{1}{M}\sum_{m}^{M}[S_m^{exp}(q)-S_m^\tau(q)]^2,
\end{equation}
where $N$ and $M$ are the numbers of points in the diagrams of the
functions $g(r)$ and $S(q)$.

\subsection{COMPUTATIONAL DETAILS}
\hfil

The effective potentials, obtained by means of the inverse method,
reproduce the structural properties of the modelled system with
good agreement with the experimental data. To check the adequacy
of the simulated systems to a real ones it is also necessary to
compare the dynamic properties of the system with their
experimental values. Therefore, the molecular dynamics (MD)
method, which enables the calculation of both structural and
dynamic properties of the system, was used in the work presented.

The MD simulations of the $3d$ liquid transition metals have been
performed under microcanonical ($NVT$) ensemble. $N$=5000
spherically symmetric particles, which interact by means of the
effective potential $\varphi(r)$, were arranged in a cubic box
with the imposed periodical boundary conditions. The size of the
box was determined from the condition that the simulated system
density is equal to the experimentally measured density of the
alloys \cite{Smithells}.

The dynamic properties such as self-diffusion coefficient $D$ and
shear viscosity $\eta$ are known to be sensitive to the specified
potential of interaction in a system. To investigate the validity
of the derived potentials the coefficients $D$ and $\eta$ have
been calculated by standard Green-Kubo (GK) formulas \cite{Allen}:
\begin{equation}\label{GK}
D=\frac{1}{3N} \int_0^{+\infty} \sum\limits_i^N<v_i(t_0)v_i(t)>dt,
\end{equation}

\begin{equation}\displaystyle\label{visk}
\eta=\frac{\rho}{k_bT}\int<\phi^{\alpha\beta}(t_0)\phi^{\alpha\beta}(t)>dt,
\end{equation}
where ${\alpha\beta}$=$xy$, $yz$, $zx$ and $\phi^{\alpha\beta}$ is defined by follow expression:
$$ \phi^{\alpha\beta}(t)=\sum\limits_{i}m_i
v_i^{\alpha}(t)v_i^{\beta}(t)-\sum\limits_{j>i}(r_{ij}^{\alpha})\frac{d\varphi(r_{ij})}{dr_{ij}^{\beta}},
$$
where $v_i(t)$ is the velocity of $i$-th particle with mass $m_i$
at the time $t$.

\subsection{STATISTICAL ANALYSIS OF THE VORONOY POLYHEDRA}
\hfil

The sensitivity of the statistical analysis of Voronoy polyhedra
to the local structure is particularly helpful for understanding
the temperature evolution of the structure of disordered systems.
The Voronoy polyhedron (VP) is defined as the sub-volume whose
interior is closer to a specific atomic vertex than to any other
vertex and may provide useful information about the local atomic
environment. The most widely applicable forms of the statistical
analysis of the VP of disorder systems is the analysis of its
metric and topological properties -- distribution of VP volumes,
face areas, topological indexes etc. However, it is more
convenient to perform the analysis by means of the non-dimensional
measure -- sphericity coefficient $K_{sph}$:
\begin{equation}\label{Ksph}
K_{sph}=\frac{36\pi V^2}{S^3}.
\end{equation}
Here, $V$ is the volume, and $S$ is the surface area of the given
polyhedron. Thus, the measure $K_{sph}$ is constructed to be unity for
regular sphere, and defines a deviation of the shape of the given
VP from regular sphere. Since each VP is the geometrical image
of the local environment of the given atom, the measure
$K_{sph}$ will characterize homogeneity of the its local environment.
The statistics of the VPs, as the distribution of the sphericity
coefficients $N(K_{sph})$, will characterize whole system and for
homogeneous systems such distribution must have symmetrical Gauss-like shape.

\section{RESULTS AND DISCUSSIONS}
\hfil

In the work the experimental structural factors $S_{exp}(q)$ of
liquid iron at $T=1833$, $1873$, $1923$, $2023$, cobalt at
$1823$, $1873$, $1923$, $2023$ and nickel at $1773$,
$1873$, $1923$ and $2023$ $K$ \cite{Waseda} were used as the
criterion functions for inverse self-consistent method. The
obtained effective pair potentials $\varphi(r)$ have reproduced the
structural properties of the simulated systems
with good agreement to the experimental ones. The deviation $\chi$
between simulated and experimental functions reaches only
$10^{-4}$ (fig.~1).

The derived potentials, presented on fig.~2 with the experimental and simulated
pair correlation functions, show long-ranged oscillations. For
metallic systems such oscillatory behavior is interpreted as
the dielectric screening of ions by conducting electrons and can be
associated with the Friedel's oscillations.

There is a hump in the main minimum of the effective potentials,
which becomes less pronounced with the temperature rise.
At motion along the $3d$ row from $Fe$ to $Ni$ the intensity of the
hump decreases as well.

It is necessary to note, that position of minimum of the derived
potentials $r_{min}$ is shifted with respect to the position of
the first maxima of RDFs $r_0$. We believe, it can evidence a
noticeable contribution of the distant neighbors in the effective
interaction. The ratio $r_{0}/r_{min}$ tends to be unity at
high temperatures and at the motion along $3d$ row, that testify that the distant contributions
weaken with temperature rise.


To investigate the validity of the derived potentials in the
present work the self-diffusion $D$ and shear viscosity $\eta$
coefficients were calculated for investigated melts.
The obtained values are presented on fig.~3 as the
function of temperature. There is reasonable agreement between
calculated and available experimental and theoretically predicted
data [18-29].

Thus, the MD configurations, obtained using the effective potentials,
can pretend on the realistic description of the structure of the  liquid
$3d$ transition metals, since reproduce both structural, and dynamic properties of the melts well.


The distributions of the sphericity coefficient $N(K_{sph})$
were calculated $Fe$, $Co$ and $Ni$ at different temperatures and presented
in fig.~4a, b, c. It is obvious that the shape of the curves $N(K_{sph})$ evolves with
the temperature rise. Their maxima are shifted towards the low
values of $K_{sph}$ that testify the distortion of the
local atomic environment in simulated liquids. At relatively low
temperatures the shoulders on the curves were obtained. This fact can be
evidence of the existence of some competitive types of the
local order. However, with the temperature rise these
shoulders disappeared and shape of the distributions $N(K_{sph})$
becomes more symmetrical, which says about more homogeneous
structure of the liquid at this temperatures.

To investigate the temperature influence on the local structure of
the alloys it is convenient to determine the most probable values
of the sphericity coefficient $K_{sph}^p$, which characterizes the
prevailing types of the local order in the structure and establish
the dependence of such values on the reduced temperature $\tau$:
\begin{equation}
\tau=\frac{T-T_m}{T_m},
\end{equation}
where $T_m$ and  $T$ are the melting and current temperatures,
respectively.

Fig.~4d demonstrates that at small values of $\tau$
the maxima of the curves $N(K_{sph})$ for all
simulated metals lie near the value $K_{sph}=0.7$. We suppose,
that this value corresponds to the common symmetry of the local
order, which is intrinsic to the dense liquids.

The slope of the dependencies $K_{sph}^{p}(\tau)$ changes at
$\tau^*$, where the distributions $N(K_{sph})$ get more asymmetrical
shape. It should note, that at $\tau>\tau^*$ the points of the
dependencies for all investigated metals can be approximated by the
common linear law (dashed line). The temperature evolution of the
distributions $N(K_{sph})$ at $\tau>\tau^*$ allows us to suppose that
the differences between the competitive types of the local order in alloys
vanish and the structure of liquid becomes more homogenous.

\section{CONCLUSIONS}
\hfil

The results of the present work allow us to suppose, that the
obtained effective potentials of liquid $3d$ transition metals at given temperatures
can reproduce their real structure, since the model adequately reflect
both structural, and dynamic properties of the melts.

The performed statistic analysis of the Voronoy polyhedrons
allow us to estimate the temperature range of inhomogeneity
$T_m-T^*$ (where $T^{*}=T_m(\tau^* +1)$ see fig.~4d). Within of this range
metallic liquid contains several competitive types of the local order.
Out of this range $T>T^*$ the structure of liquid becomes more homogenous
and the character of the local order evolution becomes common for all the investigated
metals. According to our estimation, the
boundary of the temperature range of inhomogeneity is for iron
$T^*=1873$ $K$($\tau^*=0.024$), for cobalt $T^*=1923$
$K$($\tau^*=0.061$) and for nickel $T^*=2023$ $K$($\tau^*=0.114$).
We believe, that the estimations can be useful on the choice of technological
parameters of supercooling of $3d$ metal-based alloys.

\ack The authors are highly appreciate Prof.Gelchinsky, Prof.Ladyanov and Dr.Vasin for fruitful discussions.

\newpage

\section{Figure captions}

Figure 1.
The experimental (open circles) and simulated (solid lines) structural factors $S(q)$
for liquid (a) $Fe$, (b) $Co$, (c) $Ni$ at different temperatures.
Deviation $\chi$ between the functions reaches only $10^{-4}$.


Figure 2.
The experimental (open circles) and simulated (solid lines) pair correlation functions $g(r)$ with
the effective potentials $\varphi(r)$ derived for (a) $Fe$ (b)$Co$ and (c) $Ni$ from the experimental data \cite{Waseda}.
The nearest neighbors distance $r_0$ is shifted with respect to the position of main minimum of the potential $r_{min}$.
Since the ratio $r_{0}/r_{min}$ tends to be unity at the temperature rising and at the motion along $3d$ row from $Fe$ to $Ni$,
it means the contributions of the distant neighbors to the effective interaction weaken.

Figure 3.
Values of the self-diffusion coefficient $D$ and shear
viscosity $\eta$ (open squares) calculated for the MD configurations of $Fe$, $Co$ and $Ni$
using the Green-Kubo (GK) relations, and available experimental and calculated data [1, 18-29].

Figure 4.
Distributions of the sphericity coefficient $N(K_{sph})$
calculated for liquid (a) iron, (b) cobalt, (c) nickel in the investigated temperature range;
(d) The most probable values of the sphericity coefficient $K_{sph}^{p}$ as the function of
the reduced temperature $\tau$. The slope of the dependencies $K_{sph}^{p}(\tau)$ is changed at
$\tau^*$. The behavior of $K_{sph}^{p}(\tau)$ at $\tau>\tau^*$ can be approximated
by linear low (dashed line), which is common for all metals investigated in the present work.

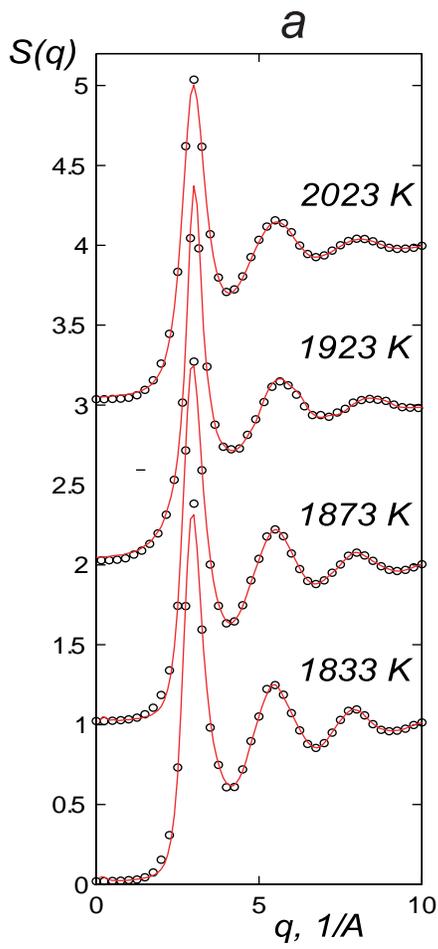
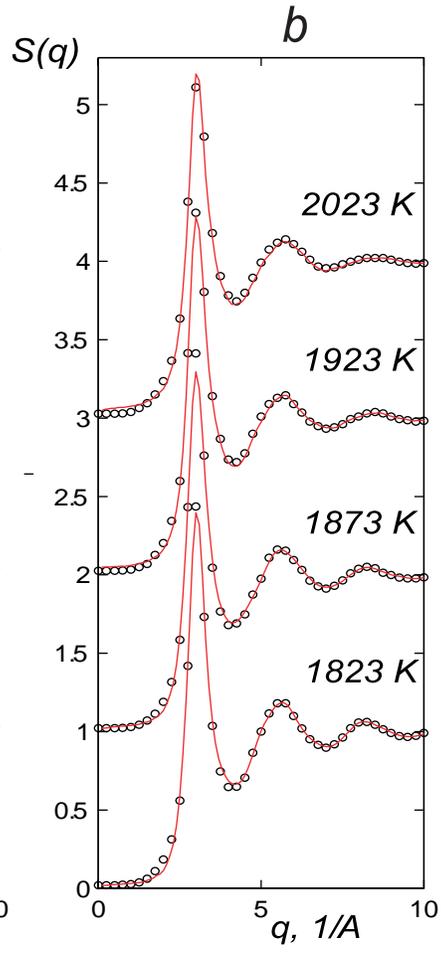
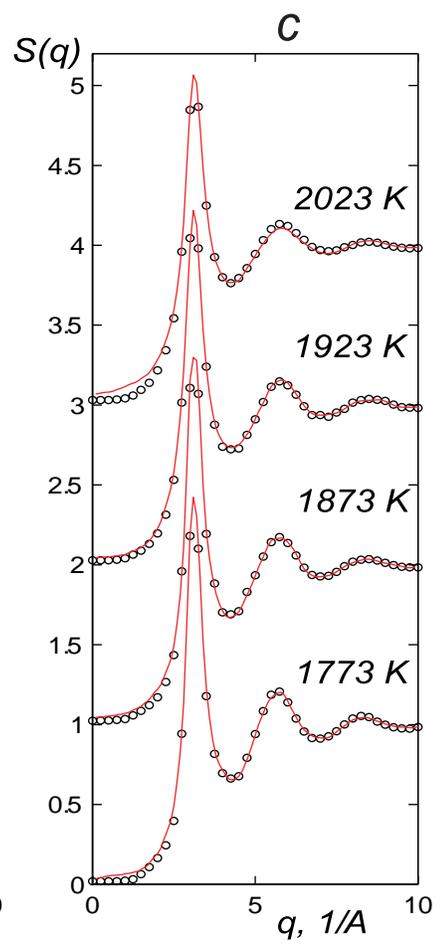
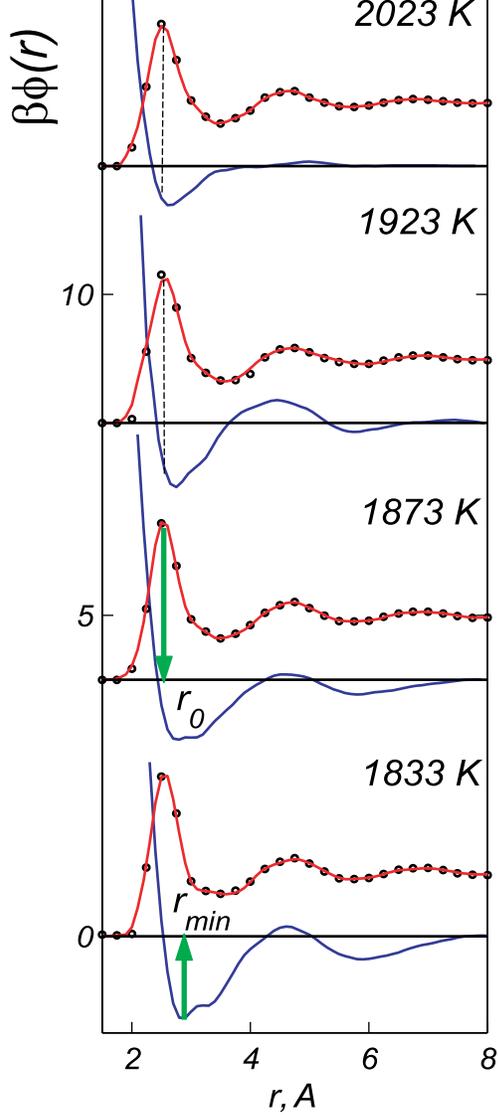
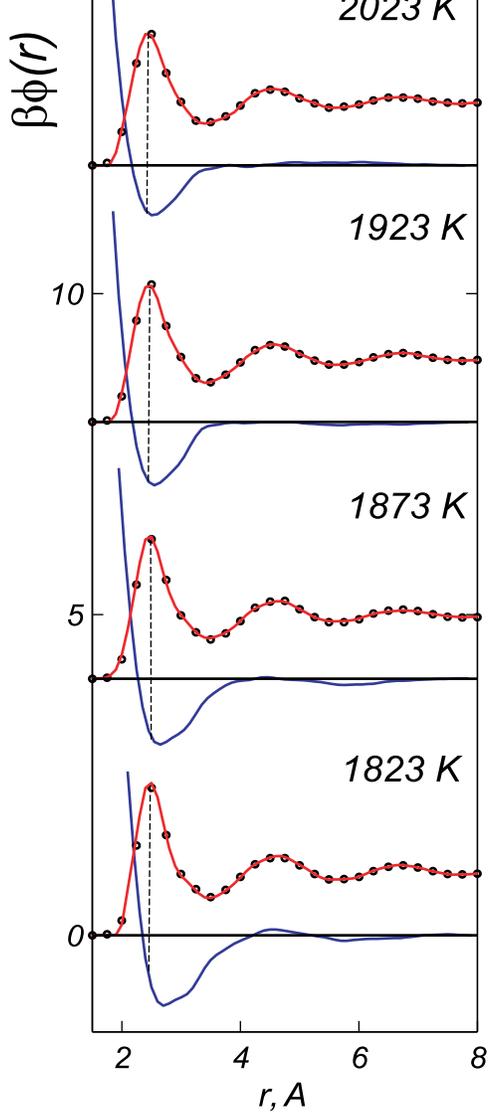
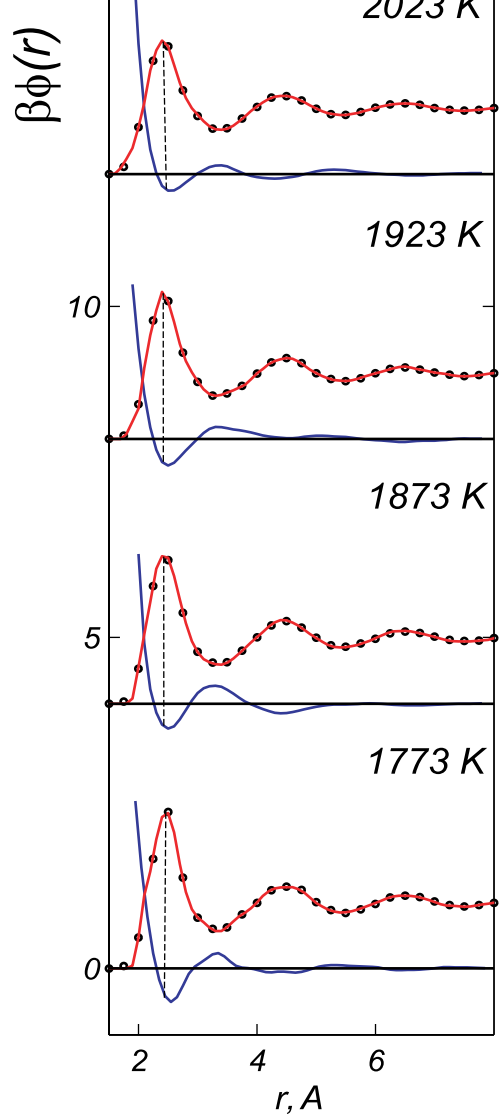
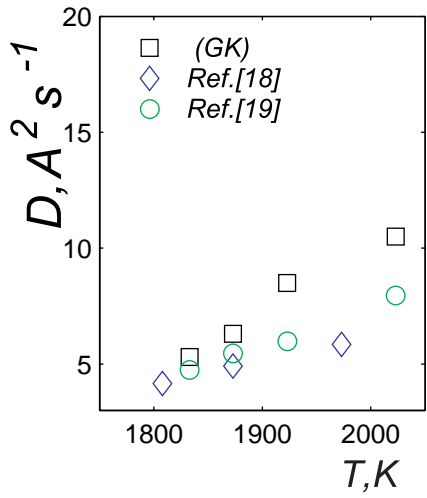
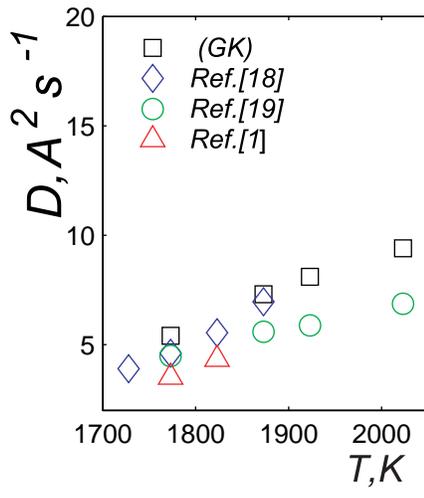
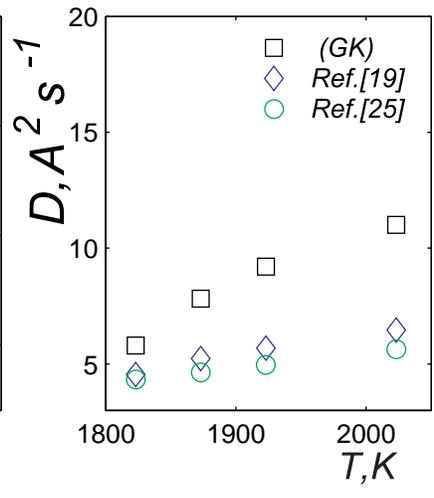
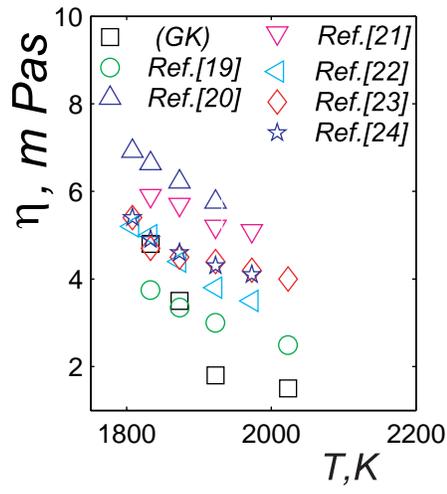
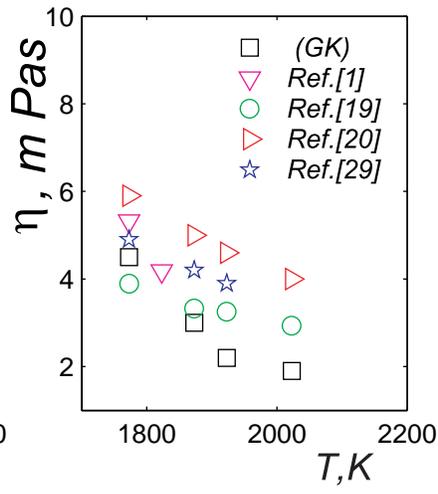
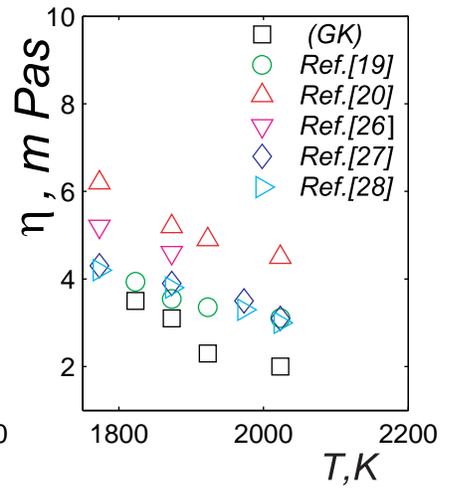
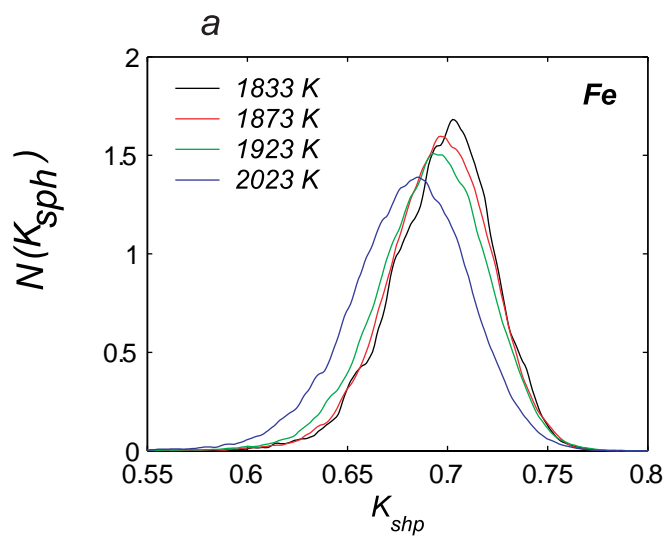
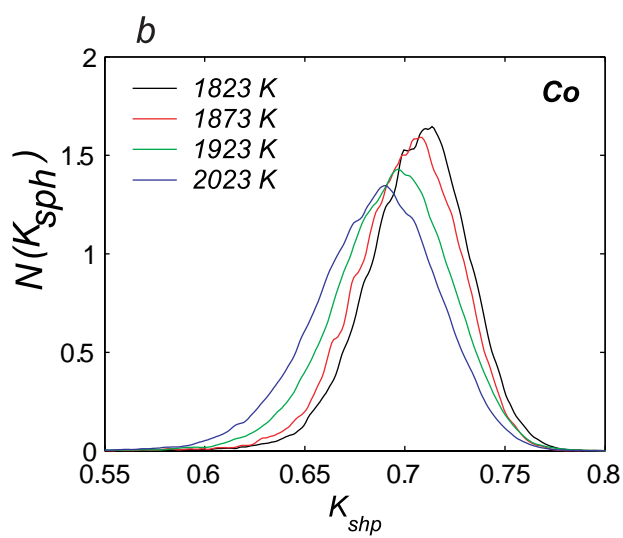
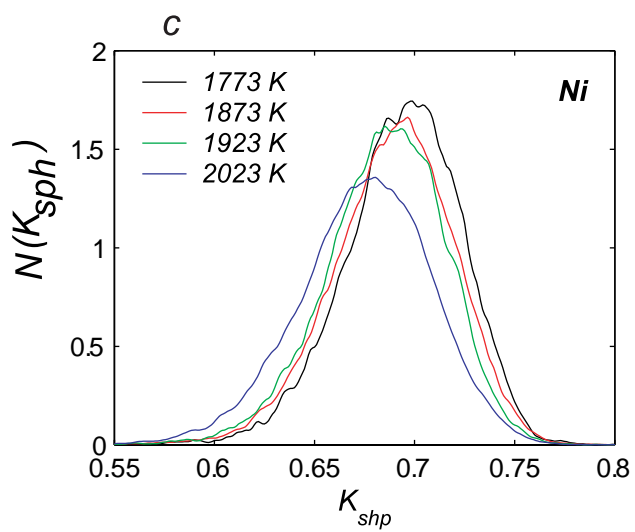
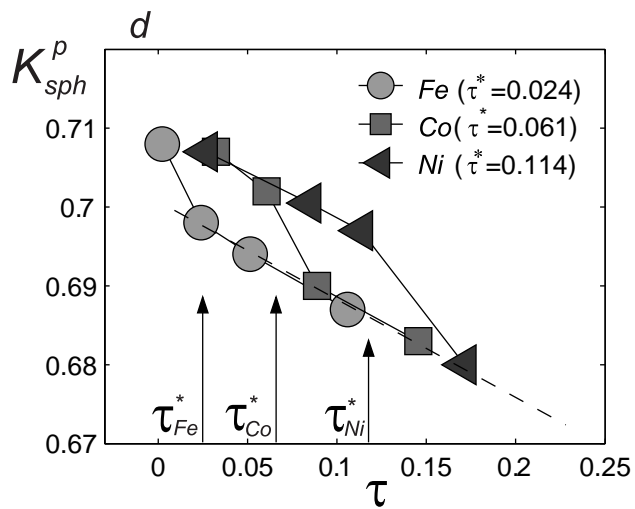
\end{document}